\documentstyle[epsfig,aps,preprint]{revtex}
\preprint{FN-IEM/2000/4}  
\tightenlines  
\draft  
\begin{document}  
\newcommand{\bd}[1]{ \mbox{\boldmath $#1$}  }
\newcommand{\xslash}[1]{\overlay{#1}{/}}
\newcommand{\sla}[1]{\xslash{#1}}
\begin{titlepage}  
\thispagestyle{empty}  
  
\begin{center}  
  
{\Large \bf Relativistic Mean Field Approximation to the Analysis of    
$^{16}O(e,e'p)^{15}N$ data at  $|Q^2|\leq 0.4$ $(GeV/c)^2$}  

\vspace{0.5cm} 
{\large J.M. Ud\'{\i}as$^{1}$, J.A. Caballero$^{2,3}$, 
E. Moya de Guerra$^{3}$, Javier R. Vignote$^{1}$, 
A. Escuderos$^3$ 
} 
 
\date{\today}
 
\vspace{.3cm} 
{$^{1}$Departamento de F\'{\i}sica At\'omica, Molecular y Nuclear, 
Universidad Complutense de Madrid,  
E-28040  Madrid, Spain}\\ 
{$^{2}$Departamento de F\'{\i}sica At\'omica, Molecular y Nuclear, 
Universidad de Sevilla,  
Apdo. 1065, E-41080 Sevilla, Spain}\\ 
{$^{3}$Instituto de Estructura de la Materia, CSIC  
Serrano 123, E-28006 Madrid, Spain}\\

\end{center} 
 
\vspace{0.5cm} 
\begin{abstract} 
We use the relativistic distorted wave impulse approximation
to analyze data  
on $^{16}O(e,e'p)^{15}N$ at $|Q^2|\leq 0.4$
(GeV/c)$^2$ that were obtained by different groups and seemed controversial.
Results for differential cross-sections, 
response functions and $A_{TL}$ asymmetry are 
discussed and 
compared to different sets of experimental data for 
proton knockout from  
$p_{1/2}$ and $p_{3/2}$ shells in $^{16}O$. We compare with a nonrelativistic
approach to better identify 
relativistic effects. The present relativistic approach is
found to accommodate most of the discrepancy between 
data from different groups,
smoothing a long standing controversy.
\vspace{0.5cm} 
 
\end{abstract} 
 
\end{titlepage} 
 
\newpage 
 
\setcounter{page}{1} 
 
\section{Introduction}

Quasielastic $(e,e'p)$ processes are a powerful tool to study bound
 nucleon properties.
 Indeed, coincidence $(e,e'p)$ measurements at quasielastic kinematics
 have provided over the years detailed information on the energies,
 momentum distributions and spectroscopic factors of bound nucleons.
 This is so because at quasielastic kinematics the $(e,e'p)$ reaction
 can be treated with confidence in the impulse approximation, {\em i.e.},
 assuming that the detected knockout proton absorbs the whole
 momentum ($q$) and energy ($\omega$) of the exchanged photon (for recent
 reviews of the subject see ref.~\cite{BGPR96} and references
  therein). Until
 recently most data were concentrated in the low missing momentum range
 $p_m\leq 300$ MeV/c, where $p_m$ is the recoil momentum of the
  residual nucleus. In the last years~\cite{Bobe94} higher $p_m$-regions
   are
 being probed at small missing energies $E_m$ to study
 further aspects of bound
 nucleon dynamics and nucleon currents.
 A substantial amount of theoretical work on $(e,e'p)$ has been carried out on the basis of
 nonrelativistic approximations to the nucleon current. This is the case 
 of the standard distorted wave impulse approximation 
(DWIA)~\cite{BGPR96}
 that uses a nonrelativistic approximation to the nucleon current operator and
 wave functions. DWIA has been successfully used over the 
years~\cite{Lapikas} to analyze $(e,e'p)$ data
 using bound and scattered proton wave functions deduced from phenomenological
 nonrelativistic
  potentials. The limits of validity of the nonrelativistic DWIA approach are
now being studied by Meucci, Giusti and Pacati~\cite{MGP01}, among others.

In past years we investigated~\cite{Udi93,Udi95,Udi96,Udi99}
nuclear responses and differential cross-sections for 
exclusive quasielastic electron scattering within the framework of 
relativistic 
mean field approximations. In the relativistic distorted wave impulse 
approximation (RDWIA)\cite{Udi93,vanorden,Jin94,Gardner}
 the one-body nucleon current,
\begin{equation}
J^{\mu}_{N}(\omega,\vec{q})=\int\/\/ d\vec{p}\/
\bar{\psi}_F
(\vec{p}+\vec{q}) \hat{J}^\mu_N(\omega,\vec{q}\/) \psi_B(\vec{p})\; ,
\label{nucc}
\end{equation}
is calculated with relativistic   $\psi_B$ and $\psi_F$
 wave functions for  initial bound
and  final outgoing  nucleons, respectively, and with relativistic nucleon
current operator, $\hat{J}^\mu_N$.
The bound state wave function is a four-spinor with well-defined parity and
angular momentum quantum numbers, and is
obtained by solving the Dirac equation with scalar-vector
(S-V) potentials determined through a Hartree procedure from a relativistic
Lagrangian with scalar and vector meson terms~\cite{HMS91}.
 The wave function for the outgoing proton is a  solution
 of the Dirac equation containing  S-V global optical potentials~\cite{CHCM93}
 for a nucleon scattered with asymptotic momentum $\vec{p}_F$.
Dirac equations for both scattered and bound wave functions are solved
in coordinate space and their solutions are then transformed to
momentum space where necessary.

Eq.~(1) sets up the scenario where differences between RDWIA and DWIA are at
play. To go from the relativistic to the nonrelativistic approach the 
one-body ($4\times 4$ matrix) current operator is first of all 
expanded in a basis
of free nucleon plane waves. This amounts to a truncation of the nucleon
propagator that ignores negative energy solutions of the free Dirac equation.
Next, a Pauli reduction~\cite{Udi95} is made to transform the current operator
into a $2\times 2$ matrix, and an expansion in powers of $(q/M)$ and/or
$(p/M)$ (where $M$ is the nucleon mass) is made~\cite{Ama96}. Finally the transition 
nucleon current is calculated
as the matrix element between bispinorial, nonrelativistic bound
($\phi_B$) and scattered ($\phi_F$) wave functions instead of the 4-component
$\psi_B$, $\psi_F$ wave functions. We then cast relativistic effects into:

i) {\em Kinematical}. These are effects due to the truncation 
of the current operator
to first, or higher order in $p/M$, $q/M$. For moderate $p/M$ values the 
relativized form proposed in \cite{Ama96} gives proper account of such effects.

ii) {\em Dynamical}. These are effects due to the differences between 
relativistic and nonrelativistic wave functions which depend not only on the 
4-spinor versus 2-spinor structure, but also on the potentials used in the
respective Dirac and Schr\"odinger equations for the bound and scattered
nucleon. 
Salient features of dynamical effects are: a) A dynamical depression of the 
upper component of the scattered nucleon wave function in the nuclear
interior, typically identified as the effect of the Darwin term coming
from the derivative of
the optical S--V potentials~\cite{Udi95}. b) A dynamical enhancement of the
lower components, mainly that of the bound nucleon wave function. 

So far, we applied successfully RDWIA to $^{208}$Pb and
$^{40}$Ca at low $|Q^2|$~\cite{Udi93,Udi95}, and to
$^{16}$O at high $|Q^2|$~\cite{Udi99,Gao2000}. 
The effect caused by the nonlocal Darwin
term for $^{40}$Ca and $^{208}$Pb cases was studied in detail 
in refs.~\cite{Udi93,Udi95}. The Darwin term causes an apparent
enhanced absorption when comparing the RDWIA 
differential cross-section to the DWIA one at moderate $p_m$ values, thus 
predicting larger spectroscopic factors~\cite{Udi93,Udi95,otros}.
For larger missing momentum 
values ($p_m/(Mc) \geq 1/3$) the lower components of 
the relativistic wave functions 
start to play a more important role, enhancing the higher momentum components 
of the nucleon wave functions. In previous work~\cite{Udi96} we found that
RDWIA calculations, 
compared to standard DWIA, tend to produce lower cross 
sections at $p_m\leq 300$ 
MeV/c and larger cross-sections at $p_m\geq 300$ MeV/c, improving 
agreement~\cite{Udi93,Udi95,Udi96} with experiment. 

The effect of the dynamical enhancement of the lower components was
studied in RPWIA in refs.~\cite{Cab98a,Cab98b}. It was also studied in RDWIA 
in ref.~\cite{Udi99} at
high $|Q^2|$. In both cases it was found to play a crucial role in the $TL$ responses.
Recent data~\cite{Gao2000} on $^{16}$O at high $|Q^2|$ seem to confirm former
RDWIA predictions.
In particular, the richness shown by the 
structure of the $A_{TL}$ asymmetry, which is different for
$p_{1/2}$ and $p_{3/2}$ shells, is only consistent with predictions 
of relativistic calculations that include the dynamical enhancement of the 
lower components of bound Dirac spinors. 
Moreover, recent data on polarization observables in $^{12}$C at
$|Q^2|\approx 0.5$ (GeV/c)$^2$ also agree nicely 
with RDWIA analysis~\cite{Woo98,Udi2000}.

For $^{16}$O
there is an important controversy in the comparison of theory to
data at low $|Q^2|$.
We refer to the data sets from 1$p$-shell proton 
knockout experiments on $^{16}$O performed at Saclay~\cite{Chin91}
and NIKHEF~\cite{Spal93,Leus94} in various kinematics
in late 80's/early 90's. These
experiments measured the cross-section as a function of missing momentum 
and, in particular Chinitz {\it et al.}~\cite{Chin91} and 
Spaltro {\it et al.}~\cite{Spal93},
extracted also the $TL$ response and $A_{TL}$ asymmetry at
$|Q^2|=0.3$ (GeV/c)$^2$ and $0.2$ (GeV/c)$^2$, respectively. 
The measurements from Chinitz {\em et al.}
were compared to relativistic~\cite{Chin91} and nonrelativistic~\cite{Spal93} 
DWIA calculations showing relatively  
small deviations from theory. On the other hand, the data of Spaltro
{\em et al.}~\cite{Spal93} were compared to results from
standard nonrelativistic DWIA calculations, and were
found to be far from theory. Using nonrelativistic 
optical potential parameters by Schwandt {\it et al.}~\cite{Schw82},
and spectroscopic factors fitted to
data in parallel kinematics, Spaltro {\em et al.}~\cite{Spal93}
found that the experimental $R^{TL}$ is enhanced
by a factor $\simeq 2.05$ 
for the $1p_{3/2}$ shell and by a factor $\simeq 1.5$ for 
the $1p_{1/2}$.
 
Though
the large discrepancy between DWIA results and experiment
found by Spaltro {\em et al.}
may in part be due to two-body 
currents, calculations of exchange current effects are still
contradictory~\cite{Sluy94,ALC99}. Hence, 
the controversy surrounding the $TL$ response and asymmetry 
data still persists.  In view of forthcoming information on
$^{16}O$ responses from experiments at Jefferson Lab in the near future, it is
important to 
reexamine these sets of data with RDWIA calculations.
We investigate whether a systematic fully relativistic analysis 
of the $(e,e'p)$ data at low $|Q^2|$
may explain the apparent discrepancies between data
from Saclay~\cite{Chin91} and NIKHEF~\cite{Spal93,Leus94}.

The paper is organized as follows: 
in Section II we summarize the basic formalism needed to describe coincidence 
electron scattering reactions, paying  special attention to 
the relativistic 
distorted wave impulse approximation (RDWIA).  
Section III contains the theoretical results obtained and 
their comparison with the experimental data. In section IV
we  present 
our conclusions. 

\section{Description of $(e,e'p)$ calculations} 

The general formalism for exclusive electron scattering reactions has been 
presented in detail in several previous papers. We refer in particular to  
refs.~\cite{BGPR96,Udi93,RD89}. 
Here we just summarize the kinematics and focus on those 
aspects that are of relevance to the points under discussion 
in this paper. As a guide to the reader we write down  
the unpolarized cross-section in  
Born approximation  
assuming plane waves for the incoming and outgoing electron (treated in the 
extreme relativistic limit), 
 
\begin{equation} 
\frac{d\sigma}{d\Omega_ed\varepsilon'd\Omega_F}= 
K\sigma_{Mott}f_{rec}\left[v_LR^L+v_TR^T+v_{TL}R^{TL}\cos\phi_F 
+v_{TT}R^{TT}\cos 2\phi_F\right] \; ,
\label{eq2}
\end{equation} 
where $\varepsilon'$ and $\Omega_e$ are the energy and solid angle  
corresponding to the scattered electron and $\Omega_F=(\theta_F,\phi_F)$ is 
the solid angle for the outgoing proton. The factor $K$ is given by 
$K=|\vec{p}_F| E_F/(2\pi)^3$, with $\vec{p}_F$ the momentum carried by the ejected proton and  
$E_F$ its energy. The term
$f_{rec}$ is the usual recoil factor $f_{rec}^{-1}=|1-(E_F/E_{A-1})
(\vec{p}_{A-1} \cdot \vec{p}_F)/|\vec{p}_F|^2|$, where $\vec{p}_{A-1}$ and
$E_{A-1}$ are the momentum and energy of the residual nucleus, respectively.
The kinematical factors are $v_L=\lambda^2$,
$v_T=\lambda/2+\tan^2\theta_e/2$, $v_{TT}=\lambda/2$,
$v_{TL}=\lambda\sqrt{\lambda+\tan^2\theta_e/2}$ with 
$\lambda=1-(\omega/|\vec{q}|)^2$, where $\omega$ and $\vec{q}$
are the energy and momentum transfer in the reaction and
$\theta_e$ the electron scattering angle.
The above  factors, that contain the dependence on  
the electron kinematics, coincide with those given
in~\cite{RD89,AD98} except 
for a factor $\sqrt{2}$ in the interference $TL$ term. We remark that
in refs.~\cite{Chin91,Spal93} a different convention for $K$ was used
(see for instance eq.~(1) of ref.~\cite{Spal93}), which
amounts to a factor $M/E_F$ of the responses presented
in this work with respect to the ones displayed 
in~\cite{Chin91,Spal93}. 

Our calculation of differential cross-sections and responses includes 
also the effect of Coulomb distortion of the incoming 
and outgoing electron waves. This breaks the simplicity of 
eq.~(\ref{eq2}), 
which is however still useful as a guide. Nevertheless, for $^{16}$O
Coulomb distortion effects in the electron wave functions
are tiny (less than 1.5\% effect on the cross-section).

The hadronic current enters only in the response functions 
$R^\alpha$, $\alpha=L,T,TL,TT$ where $L$ and $T$ denote the  
longitudinal and transverse 
projections of the nuclear current with respect to the
momentum transfer 
$\vec{q}$, respectively. Note that the response functions 
can be separated 
by performing measurements with different kinematical 
factors $v_\alpha$ and/or 
values of the azimuthal angle $\phi_F$, while keeping the
momentum and energy 
transfer constant.
 The response
$R^{TL}$ is obtained from differential cross-sections at $\phi_F=0^o$
and $180^o$, both in theory and in experiment.
Experimental data for the cross-section are often
presented in terms of {\em reduced cross-sections} or
{\em effective momentum distributions} $\rho(\vec{p}_m)$, 
obtained by integrating
over a particular missing energy peak the differential 
cross-section divided by
$K(2\pi)^3\sigma_{ep}$. Thus $\rho(\vec{p}_m)$ is defined by,
\begin{equation}
\rho(\vec{p}_m)=\int_{\Delta E_m} 
\left( \frac{d\sigma}{d\Omega_ed\varepsilon'd\Omega_FdE_F}/(K(2\pi)^3\sigma_{ep})
\right) dE_m \; .
\label{reduced}
\end{equation}

The free electron-proton cross-section, $\sigma_{ep}$,
is usually taken as $\sigma_{CC1}$ of de Forest~\cite{Fore83}.
One must be aware that the cross-section given in eq.~(\ref{eq2}) 
has a strong dependence in the kinematical variables via $K$ and
$\sigma_{ep}$ which is removed in the reduced cross-section.
For instance, at the kinematics
of the experiment of Chinitz {\em et al.}~\cite{Chin91}
($T_F=160$ MeV, $|Q^2|=0.3$ 
(MeV/c)$^2$ and $\varepsilon_{beam}=580$ MeV)
a small variation of 5 MeV in $T_F$ and $\omega$ (keeping $E_m$ and
$p_m$ constant), may change the cross-section by as much 
as 7\% and the reduced cross-section by less than 2\%.
In order to minimize kinematical dependences, it is safer to determine 
spectroscopic factors by scaling the theory to data on reduced cross
sections rather than to data on cross-sections. This is so because
experimentally, a folding and average of the cross-sections,
responses and/or reduced cross-sections is performed
over the experimental acceptance, and central values for the kinematical
variables are quoted. Theoretical calculations are done for
the quoted central values. Due to this, it is not unusual that 
spectroscopic factors 
may depend on whether one chooses to set the scale
by comparing to reduced cross-sections
or to differential cross-sections, or even to separate responses.
In this work we first derive the spectroscopic factor
($S_\alpha$) from the reduced cross-section data. Then we use
this same factor to compare to data for the individual responses.
In this way the
analysis of $R^{TL}$ and other responses is
more consistent and meaningful.

Another quantity also obtained by the experimentalists and discussed in
next section is the asymmetry $A_{TL}$ given by
\begin{equation}
A_{TL}=
\frac{\sigma(\phi_F=180^o)-\sigma(\phi_F=0^o)}
{\sigma(\phi_F=180^o)+\sigma(\phi_F=0^o)} \; . 
\end{equation}
One can see from eq.~(2) that
this observable is closely related to $R^{TL}$, with the advantage that
it is free from the scale factor ambiguity.

\subsection{Relativistic Distorted Wave Impulse Approximation (RDWIA)} 
 
In RDWIA the process is described~\cite{Udi93} in terms of the
one-body nucleon current  given 
in eq.~(\ref{nucc}). The relativistic
bound nucleon wave function, $\psi_B$, is a four-spinor with well defined  
angular momentum quantum numbers $\kappa$, $\mu$, corresponding to the 
shell under consideration. In coordinate space it is given by: 
\begin{equation} 
\psi^{\mu}_{\kappa}(\vec{r}) 
=\left(\begin{array} 
{@{\hspace{0pt}}c@{\hspace{0pt}}} 
g_{\kappa}(r) 
\phi^{\mu}_\kappa(\hat{r}) \\ 
i f_{\kappa}(r) \phi^{\mu}_{-\kappa}(\hat{r})\end{array}\right)\; , 
\label{eq3} 
\end{equation} 
which is eigenstate of total angular momentum with eigenvalue 
$j=|\kappa|-1/2$,  
 
\begin{equation} 
\phi_\kappa^\mu(\hat{r})=\sum_{m,\sigma}\langle l m \frac{1}{2} 
\sigma|j\mu \rangle
Y_{lm}(\hat{r})\chi^{\frac{1}{2}}_\sigma 
\label{eq4}
\end{equation} 
with  
$l=\kappa$ if $\kappa>0$ and $l=-\kappa-1$ if $\kappa<0$.  
The functions $f_{\kappa},g_{\kappa}$ 
satisfy the usual coupled linear differential 
equations~\cite{Udi93,Ube71,SW86}. 
 
The wave function for the outgoing proton, $\psi_F$, is a scattering solution 
of the Dirac equation, which includes S--V global optical potentials. 
This wave function is obtained as a partial wave 
expansion in configuration space~\cite{Udi93,Udi95}: 
\begin{equation} 
\psi_F(\vec{r})=4 \pi \sqrt{\frac{E_F+M}{2E_F V}} 
\displaystyle \sum_{\kappa,\mu,m} 
e^{-i\delta^{*}_{\kappa}} 
i^{l} \langle l \ m \ \frac{1}{2} \ \sigma_F | j \ \mu \rangle \nonumber \\
Y_{lm}^{*}(\hat{P}_F)\psi_{\kappa}^{\mu}(\vec{r}) \; , 
\label{eq5} 
\end{equation} 
where $\psi_\kappa^\mu(\vec{r})$ are four-spinors 
of the same form as that in eq.~(\ref{eq3}). 
The phase-shifts and radial functions are complex because of the 
complex potential. 
 
The choice of the current operator $\hat{J}^\mu$ is to some extent 
arbitrary (see discussion in refs.~\cite{Udi93,Cab98a,CDP93}).
Here we consider the two most popular choices denoted as CC1 and  
CC2~\cite{Fore83}: 
\begin{eqnarray} 
\hat{J}^\mu_{CC1} &=& (F_1+F_2)\gamma^\mu-\frac{F_2}{2M}( 
\overline{P}+P_F)^\mu \label{eq6} \\ 
\hat{J}^\mu_{CC2} &=& F_1\gamma^\mu +i\frac{F_2}{2M}\sigma^{\mu\nu}Q_\nu\; ,
\label{eq7}
\end{eqnarray} 
where $F_1$ and $F_2$ are the nucleon form factors related in the usual 
way~\cite{BD64} to the electric and magnetic 
Sachs form factors of the dipole 
form. The variable $\overline{P}$ in eq.~(\ref{eq6}) 
is the four-momentum of the 
initial nucleon for on-shell kinematics, {\em i.e.}, $\overline{P}^\mu= 
(E(p),\vec{p})$ ($E(p)=\sqrt{\vec{p}^2+M^2}$ and $\vec{p}= 
\vec{p}_F-\vec{q}$). 

Thus
the evaluation of the one-body current matrix element involves 
the use of $4\times 4$-operators and 4-spinors with negative energy 
components. This is at variance with the nonrelativistic (DWIA) approximation
where a truncated current operator is used~\cite{DWEEPY} and matrix elements
are evaluated between bispinorial wave functions ($\phi_B,\phi_F$).
Therefore in the discussion of results in next sections we shall refer to 
relativistic kinematical effects --that have to do with the 
differences due to the use of the
complete relativistic current operator instead of the truncated one-- and
to relativistic dynamical effects. 
A way to 
 fully incorporate the
kinematical relativistic effects was suggested in ref.~\cite{Udi95,Ama96},
and studied in detail in ref.~\cite{Jeschonnek} for the reaction $^2H(e,e'p)$.

\subsection{Remarks on relativistic dynamical effects}

As mentioned in the introduction the 
dynamical effects come from the differences between relativistic and 
nonrelativistic potentials and wave functions. 
In ref.~\cite{Udi95} we discussed in detail effects
on reduced cross-sections for $^{208}$Pb in parallel kinematics
due to differences between the upper components of the four-spinors
$\psi_B$, $\psi_F$ which are Dirac solutions with S-V potentials and the
bispinors $\phi_B$, $\phi_F$ which are Schr\"odinger solutions with standard
(Wood-Saxon type) potentials for bound and scattered nucleons.

To illustrate the meaning of this effect we recall that the Dirac equation with
S-V potentials
\begin{equation}
(\tilde{E}\gamma_0-\vec{p}\cdot\vec{\gamma}-\tilde{M})\psi =0
\end{equation}
with
\begin{eqnarray}
\tilde{E}&=&E-V(r) \\
\tilde{M}&=&M-S(r) \\
\psi&=&\left(\begin{array} 
{@{\hspace{0pt}}c@{\hspace{0pt}}} 
\psi_{up} \\ 
\psi_{down}\end{array}\right)\; ,
\end{eqnarray}
can be written either as a system of coupled linear differential equations
for $\psi_{up}$, $\psi_{down}$, or as a second order differential Schr\"odinger
like-equation for $\psi_{up}$ containing also a first order derivative term
(the Darwin term). Furthermore, using the transformation
\begin{equation}
\psi_{up}(r)=K(r)\phi(r),
\end{equation}
the non-local (Darwin) term can be eliminated
to obtain a more standard Schr\"odinger equation with second derivatives only
\begin{equation}
\left[\frac{-\vec{\nabla}^2}{2M}-U_{DEB}\right]\phi(\vec{r})=
\frac{(E^2-M^2)}{2M}\phi(\vec{r})
\end{equation}
with
$U_{DEB}$ the Dirac equivalent 
potential~\cite{Udi95} with central and spin-orbit terms
\begin{eqnarray}
U_{DEB}&=&V_C+V_{SO}\,\,\vec{\sigma}\cdot\vec{\ell} \nonumber \\
V_C &=&\frac{1}{2M}\left[V^2-2EV-S^2+2MS+V_D\right]\nonumber \\
V_D &=&\frac{1}{rA}\frac{\partial A}{\partial r}+\frac{1}{2A}
      \frac{\partial^2A}{\partial r^2}-\frac{3}{4A^2}
      \left(\frac{\partial A}{\partial r}\right)^2 \nonumber \\
V_{SO} &=&\frac{1}{2M}\frac{1}{rA}\frac{\partial A}{\partial r} \nonumber \\
A(r) &=&\frac{\tilde{E}+\tilde{M}}{E+M}=K^2(r) \; .
\end{eqnarray}

The factor $K(r)$ relating the upper component of the Dirac solution
($\psi_{up}$) to the solution of the equivalent Schr\"odinger equation
($\phi(r)$) is called the Darwin factor.  As it will be shown in next
section, $K(r)$ produces a depletion of the outgoing wave 
function in the nuclear
interior \cite{Rawitscher,Cannata}.

Another dynamical relativistic effect is that coming from the non-zero overlap
with Dirac sea of the Dirac solutions with S-V potentials. 
The $\psi_B$, $\psi_F$ 
wave functions have the general structure
\begin{equation} 
\psi(\vec{p}) 
=\psi^{(+)}(\vec p)+\psi^{(-)}(\vec p)\; ,
\label{eq10} 
\end{equation}
where $\psi^{(+)}$ and $\psi^{(-)}$ are the projections on the positive and
negative energy solutions of the Dirac equation for free particles:
\begin{eqnarray} 
\psi^{(+)}(\vec{p})&=&\sum_su_s(\vec p)\overline{u}_s(\vec p)\psi(\vec p)=
\Lambda_{(+)}(\vec{p})\psi(\vec{p}) \label{eq11a} \\
\psi^{(-)}(\vec{p})&=&-\sum_s v_s(\vec p)\overline{v}_s(\vec p)\psi(\vec p)=
\Lambda_{(-)}(\vec{p})\psi(\vec{p})\; , \label{eq11b}
\end{eqnarray}
where we use the notation and conventions of Bjorken and Drell~\cite{BD64},
so that the positive and negative energy projectors are
\begin{equation}
\Lambda_{(\pm)}(\vec{p})=\frac{M\pm\sla{\overline{P}}}{2M}
\label{projector}
\end{equation} 
with $\overline{P}_\mu=(\overline{E},\vec{p})$ and 
$\overline{E}=\sqrt{p^2+M^2}$.

The positive and negative energy components of $\psi$ can also be written as
\begin{eqnarray} 
\psi^{(+)}(\vec{p}) 
&=&\left(\begin{array} 
{@{\hspace{0pt}}c@{\hspace{0pt}}} 
\psi_{up}^{(+)}(\vec p) \\ 
\psi_{down}^{(+)}(\vec p)\end{array}\right)
=\sum_s\tilde{\alpha}(\vec p,s)u_s(\vec p) \label{eq14a} \\
\psi^{(-)}(\vec{p}) 
&=&\left(\begin{array} 
{@{\hspace{0pt}}c@{\hspace{0pt}}} 
\psi_{up}^{(-)}(\vec p) \\ 
\psi_{down}^{(-)}(\vec p)\end{array}\right)
=\sum_s\tilde{\beta}(\vec p,s)v_s(\vec p)\; . \label{eq14b}
\end{eqnarray}

Eqs.~(\ref{eq14a},\ref{eq14b}) make it more transparent
what are the new ingredients
of the relativistic calculation. In particular the difference between
$\psi_{down}$ and $\psi_{down}^{(+)}$ is what we call the
dynamical enhancement of the down component. Explicit expressions and
figures showing $\tilde{\alpha}(p)$ and $\tilde{\beta}(p)$ for
several orbitals can be found in ref.~\cite{Cab98b}. Here we just mention that
the dynamical enhancement of the down
component is proportional to the nonzero Dirac sea overlap,
$\tilde{\beta}(p)$, and that
though it is small, it was found to play an
important role in the $TL$ response function in the RPWIA calculations
presented in ref.~\cite{Cab98b}, and in the RDWIA calculations at high $|Q^2|$
reported in refs.~\cite{Udi99,Gao2000}. Its role for the present RDWIA calculations
is discussed in next section.
A way to define an effective 2$\times$ 2 current operator that includes these 
dynamical relativistic effects was introduced in ref.~\cite{Heda95}.

\subsection{Projected calculation}

The sensitivity of the different scattering observables to the negative 
energy components can be analyzed by constructing properly normalized 
4-spinors of the form in eq.~(\ref{eq11a}).
Then, one can compare the results obtained using  
the fully relativistic amplitude given in eq.~(\ref{nucc}) with those 
obtained when the negative energy components are projected out. This is
done 
when the nucleon current is calculated as 
\begin{equation} 
J^\mu_{(++)}(\omega,\vec{q})=\int d\vec{p} 
\overline{\psi}_F^{(+)}(\vec{p}+\vec{q})\hat{J}^\mu(\omega,\vec{q}) 
\psi_B^{(+)}(\vec{p})\; , 
\label{eq9}
\end{equation} 
where $\psi_B^{(+)}$ ($\psi_F^{(+)}$) is the positive-energy projection 
of $\psi_B$ ($\psi_F$), {\it i.e.,}
\begin{equation}
\psi_B^{(+)}(\vec{p})=\Lambda_{(+)}(\vec{p})\psi_B(\vec{p}), \,\,\,\, 
\psi_F^{(+)}(\vec{p}+\vec{q})=
\Lambda_{(+)}(\vec{p}+\vec{q})\psi_F(\vec{p}+\vec{q})\; .
\label{eq10a}
\end{equation} 

The dynamical enhancement of the lower components is contained in the 
current of eq.~(\ref{nucc}), 
but not in eq.~(\ref{eq9}). It is important to 
realize that the positive-energy projectors inserted in eq.~(\ref{eq9}) 
depend on the integration variable $\vec{p}$. One could also neglect this 
$\vec{p}$-dependence by using projection operators corresponding to 
asymptotic values of the momenta, {\it i.e.}, 
projectors acting on $\psi_F$ and 
$\psi_B$ respectively, with $P_F^\mu=(E_F,\vec{p}_F)$, 
$P_B^\mu=P_F^\mu-\overline{Q}^\mu$ the asymptotic four-momentum 
of the outgoing and bound nucleon respectively, with $\overline{Q}^\mu= 
(\overline{\omega},\vec{q})$ and $\overline{\omega}=E_F- 
\sqrt{(\vec{p}_F-\vec{q})^2+M^2}$. We refer to this approach as {\sl 
asymptotic projection}. This latter projection
is almost equivalent to  ``EMA-noSV'' procedure employed in 
ref.~\cite{Kel97}, in which the 4-spinors
used have upper components identical to the upper
components of the Dirac equation solutions, but the lower
components are obtained with an additional approximation,
the effective momentum approach (EMA).
Although  EMA-noSV  approach also neglects 
the enhancement of the lower components, it is not at all 
equivalent to the exact projection method in eqs.~(\ref{eq9},\ref{eq10a}).
The EMA-noSV approach
computes the nucleon current with four-spinors that have  the
same structure than the ones encountered in the scattering
of free nucleons, because it enforces the relationship
between upper and lower components to be driven by the
asymptotic value of the  momenta at the nucleon vertex.
In particular, the Gordon transformation
is exact for EMA-noSV approach. Therefore, CC1 and CC2 
operators would lead to identical results within  EMA-noSV,
provided  the same
choices for the off-shell values of $\omega$, $E$, $E_F$, $\vec{p}$ and
$\vec{p}_F$ are made.
This would be a strong prerequisite to a factorized calculation, though
still not a sufficient condition. In order to keep the
drawings in section III clear enough,
we do not show in the figures  the results obtained
within EMA-noSV,
but we shall comment how this approach compares with the fully 
relativistic and/or the projected one.

\section{Results and Discussion}

In this work we consider three data sets for nucleon knockout from
$p_{1/2}$ and $p_{3/2}$ shells in $^{16}$O that correspond to
kinematical conditions of three different experiments. We summarize them as
follows:

{\bf Set $(a)$} corresponds to the experiment of Leuschner {\em et al.} 
at the Medium Energy Accelerator
(MEA) at NIKHEF-K~\cite{Leus94}. 
The coincidence reaction $^{16}O(e,e'p)^{15}N$
was analyzed in quasielastic parallel kinematics at three different beam
energies: 304, 456 and 521 MeV. The total kinetic energy of the outgoing
proton was around 90 MeV. The spectral function of $^{16}$O was
measured in the range, $0<E_m<40$ MeV and $-180<p_m<270$ MeV/c,
where $E_m$ and
$p_m$ are the missing energy and missing momentum, respectively.

{\bf Set $(b)$} corresponds to the experiment performed at the Saclay Linear
Accelerator by Chinitz {\em et al.}~\cite{Chin91}. The kinematical
setup was constant $|\vec{q}|-\omega$ kinematics. The electron
beam energy was $\varepsilon=580$ MeV, the outgoing proton kinetic energy
$T_F=160$ MeV, and the transfer momentum and energy: $|\vec{q}|=570$ MeV/c
and $\omega=170$ MeV ($|Q^2|=0.3$ (GeV/c)$^2$). The missing energy
resolution was 1.3 MeV, which made not possible to resolve the
($5/2^+$,$1/2^+$) doublet at an excitation energy $E_x=5.3$ MeV in
$^{15}N$ from the $3/2^-$ state at $E_x=6.3$ MeV.

{\bf Set $(c)$}, also in $|\vec{q}|-\omega$ constant kinematics, 
was obtained by Spaltro {\em et al.}~\cite{Spal93}
with the two high-resolution magnetic
spectrometers at the medium-energy electron accelerator MEA of NIKHEF-K.
Data were measured at
momentum and energy-transfer values centered
at $(\omega,|\vec{q}|)=(90$ MeV, $460$ MeV/c), 
{\it i.e.}, close to the center of the
quasielastic peak at $|Q^2|\simeq 0.2$ (GeV/c)$^{2}$. The experiment covered a
missing momentum range from 30 to 190 MeV/c. The missing energy
resolution was about 180 keV, which made it possible to resolve the
$(5/2^+,1/2^+)$ doublet from the $3/2^-$ state.

Next we discuss our results for spectroscopic factors, reduced cross-sections
and responses corresponding to these sets of data and kinematical conditions.

In subsection III.A we deduce spectroscopic factors from reduced cross
sections, that are then used in subsection III.B for response functions.
Subsection III.A discusses also results corresponding to different relativistic
S-V potentials. In our previous work on $^{40}$Ca and $^{208}$Pb we found
that spectroscopic factors were larger than the ones obtained with the
nonrelativistic analyses and were very stable when different parameterizations
of the S-V potentials for bound (HS, NLSH)~\cite{Udi93,Udi96} and scattered
(EDAI, EDAD1, EDAD2, EDAD3)~\cite{Udi95} protons were used. 
We shall see that the case
of $^{16}$O that we examine here is different in several respects.

\subsection{Reduced Cross Section and Spectroscopic Factors}

Let us first discuss the comparison of theory and experiment on
reduced cross-sections for set $(a)$ 
(Leuschner {\it et al.}~\cite{Leus94}). We recall
that because of  parallel kinematics, for this set the only response
functions that contribute to the cross-sections are $R^L$ and $R^T$.
Fig.~1 shows the reduced cross-section for $p_{1/2}$ 
and $p_{3/2}$ shells. The sign of
$p_m$ refers to the projection of the initial nucleon momentum along the 
direction of the transfer momentum $\vec{q}$. It is defined to be
positive for 
$|\vec{q}|<|\vec{p}_F|$ and negative for $|\vec{q}| >|\vec{p}_F|$. 
Fully relativistic calculations using the CC1  and CC2  
current operators (RCC1, RCC2) are shown by thin and thick 
lines respectively. Throughout this paper we use the Coulomb gauge. 
The Landau gauge produces similar results. Gauge ambiguities \cite{Naus90}
are rather small for the fully relativistic 
results in these two gauges~\cite{Cab98a,Cab98b}.

Spectroscopic factors for each of the two shells are evaluated by scaling 
 theoretical calculations to  experimental data.
They are listed in Table~I for different choices
of wave functions and current operators. In this table we also quote
the statistical error within parenthesis and the $\chi^2$ values 
per degree of freedom.

Results on the right panel in Fig.~1
correspond to bound state wave function calculated using
the parameters of the set NLSH~\cite{Shar93}.
Results with the older HS set~\cite{HMS91,SW86}, as well as with the
newest NL3 one~\cite{LKR97} are similar.
For the scattered proton wave function we use the energy-dependent
$A$-independent potential derived by Clark {\it et al.}~\cite{CHCM93} for
$^{16}$O (EDAI-O). 
Two things are striking in these results that are at variance with the 
situation we met in previous works on $^{40}$Ca and 
$^{208}$Pb~\cite{Udi93,Udi95}:
\begin{enumerate}
\item There are clear deviations in the shape of theoretical and
experimental effective momentum distributions in the right hand side panel
of Fig.~1. Actually, the NLSH wave functions have smaller (larger) r.m.s.
radii in r-space (p-space) than what is shown experimentally.
\item The spectroscopic factors are small, of the same order or even
smaller than nonrelativistic ones when Perey factor is included in the
latter. As seen in Table~I the spectroscopic factors increase when global
$A$-dependent type (EDAD-1,-2) potentials are used instead of 
the $A$-independent
potential fitted to $^{16}$O (EDAI-O). Moreover, the $\chi^2$ values are
large for NLSH bound wave functions independently of the optical potential
used.
\end{enumerate}

We have verified that all EDAD-type calculations (EDAD-1,-2,-3) give similar
results on reduced cross-sections and responses. Compared to EDAI-O they give
about $15-20\%$ smaller reduced cross-sections with almost identical shapes. 
Consequently,
EDAD-1,-2,-3 spectroscopic factors are $15-20\%$ larger than EDAI-O ones
but the $\chi^2$ values are analogous (see Table~I). Why this is  different
from the cases we analyzed in refs.~\cite{Udi93,Udi95} can be easily understood
from Fig.~2. In this figure we compare the relativistic central potentials ($S, V$) 
and Darwin factors ($K$) corresponding to EDAI-O and to EDAD-1,-2 
optical potentials
for $^{208}$Pb (right panels) and for $^{16}$O (left panels). We can see that
in the case of $^{16}$O, EDAD-1,-2 potentials produce a deeper $K(r)$, 
{\it i.e.,} a larger reduction of the scattered wave in the nuclear interior 
than EDAI-O potential --also, $V_C$ is somewhat more absorptive--, 
while in the case of $^{208}$Pb both are about the
same. Consequently, at the energies considered, EDAD-1,-2 potential lead to
larger spectroscopic factors than EDAI-O in $^{16}$O, while the two potentials
lead to similar spectroscopic factors in $^{208}$Pb. The same is true for
EDAD-3 and other versions of the relativistic EDAD potentials.

To have a more conclusive determination of $^{16}$O spectroscopic factors
one would need to constrain the optical potential choice by means of inelastic
($p,p'$) data, in addition to the elastic ones \cite{KellyEEI}. 
But this is not available for
the small knock-out proton energies ($\sim 90$ MeV) considered here nor in a
fully relativistic framework.

The large $\chi^2$ values in the left part of Table~I for all the optical
potentials have to do with the fact that the data do not follow the shape
of theoretical reduced cross-sections in the right panel of Fig.~1. A similar
problem has been found for data sets ($b$) and ($c$) where the quality of
NLSH fits is even worse. This, after all, is not surprising because the 
standard Lagrangians, like NLSH, are fitted to bulk properties of a few
heavy nuclei, and one may expect that the predicted r.m.s. radii of $^{16}$O
orbitals differ somewhat from experiment. Unfortunately, as seen in Table~I
this produces large uncertainties in spectroscopic factors. To solve this
problem we may adjust the parameters of the relativistic potentials (or
Lagrangian) so as to obtain the correct values of the single particle energies
and r.m.s. radii for the orbitals considered, in an analogous way to what is 
usually done in nonrelativistic analyses of $(e,e'p)$ data. This is what we do next.

Compared to data sets $(b)$ and $(c)$, data set $(a)$ has many
more data points extending over a larger $p_m$ range.
Therefore, this data set can be used much more reliably to
determine  {\em simultaneously} spectroscopic
factors and r.m.s.r. values. We have then adopted the following
strategy: First, we use data set $(a)$ to slightly tune the
parameters of the NLSH potential so as to reproduce the experimental 
binding energies and r.m.s.r. 
values of the $p_{1/2}$ and $p_{3/2}$
orbitals in $^{16}$O, closely resembling the standard nonrelativistic
procedure. We denote by NLSH-P the new relativistic
potentials and wave functions (see Table~II). These new relativistic
 wave functions are 
then used to make predictions for the kinematical
conditions of   data sets $(b)$ and $(c)$.

The NLSH-P wave functions  are obtained
by changing the parameters of the NLSH Lagrangian  so that 
the radii and depth of the S and V potential wells derived from 
the Lagrangian are modified in the same proportion. 
The negative energy content of the 
resulting bound state wave function is barely changed by this procedure.
The rescaling of the depth size and  radii of the NLSH-P wells is
within 10\% of the initial NLSH ones.
The improvement obtained in the description of the shapes and quality
of the fits is clearly visible in Fig.~1 and Table~I.

The role played by relativistic dynamical effects is also 
analyzed from the results presented in Fig.~1 and Table~I.
Each curve in Fig.~1 is scaled by the corresponding spectroscopic factors
in Table~I. The reduced cross-sections 
evaluated after projecting the bound and scattered proton wave functions over
positive energy states (see eq.~(\ref{eq9})) 
are shown by thin-dashed (PCC1) and 
thick-dashed (PCC2)
lines. Note that the difference between PCC1 and PCC2 
results is very small because the so-called Gordon ambiguities are reduced 
after projection~\cite{Cab98a,Cab98b}. 
The results obtained using the asymptotic
values of the momenta in the projection operator
as described in  section II.B, are almost identical to
the PCC2 results and thus are not shown here.
Once the global scale factor is taken into account, all
the calculations predict a very similar behavior, what  indicates
that, aside from the Darwin term,
the effect of relativistic dynamics in the reduced cross-sections is
not important in parallel kinematics at low values of $|Q^2|$. This agrees with
a recent work by Giusti and collaborators~\cite{MGP01}.
This observation also agrees with results of
some previous works~\cite{Udi99,Cab98a,Cab98b}
where we saw that the dynamical enhancement of
the lower component makes an  important effect in the cross-section
mainly at high missing
momentum values and/or in the $R^{TL}$ response function 
(which {\em does
not contribute in parallel kinematics}), whereas its influence on
$R^L$ and $R^T$ is quite modest.

Comparing the fully relativistic results with NLSH-P wave functions for
CC1 and CC2, one observes that the differences are at most of the order
of $\sim 8\%$.
In the case of the projected calculation, we  note
that the spectroscopic factors are
slightly larger than those corresponding to a fully relativistic calculation.
This is due to the enhancement of the lower components of the
wave functions which is not contained in the projected approximations.
Their effect is negligible for CC2 operator and is enhanced
by the CC1 choice. 
We recall that another
dynamical relativistic effect, namely the Darwin term, is contained in
all the figures and tables shown here. For EDAI-O optical potential this
effect amounts to a 10\% reduction of
the reduced cross-section in $^{16}$O for the kinematics 
discussed in this work. This is comparable to the effect of the
Perey factor that was included in nonrelativistic DWEEPY 
calculations~\cite{Spal93,Leus94} 
 while for EDAD-optical potentials this amounts to
a 20$\%$ reduction.

One thus expects  the spectroscopic factors listed in Table~I 
for EDAI-O in the projected case to be similar to those
obtained from fits with  standard nonrelativistic DWIA calculations including
Perey factors. In this
last case the extracted factors for various choices of  optical potentials are
$0.60\leq S_\alpha \leq 0.65$ for $p_{1/2}$, and
$0.49\leq S_\alpha \leq 0.60$ for $p_{3/2}$~\cite{Leus94}, which 
are also roughly in agreement with those in Table~I for NLSH-P and
EDAD-type potentials.

A smaller spectroscopic factor  is expected for the $p_{3/2}$ shell
than for the $p_{1/2}$, because the $p_{3/2}$ strength is known
to be fragmented
into three states: the state considered here at $E_m=18.4$ MeV, and two
weaker peaks at around 22.0 and 22.7 MeV. 
According to ~\cite{Leus94,Chin-PhD}
the two higher lying peaks would contain about 10\% of the 
total $p_{3/2}$ strength.
The spectroscopic factors determined from data set $(a)$ indicate that,
taking this extra  10\% contribution into account, 
there is similar $3/2^-$
and $1/2^-$ spectroscopic strength.

In what follows we use the new bound state wave functions (NLSH-P) 
to make predictions for comparison to the other data sets $(b)$ and $(c)$.
We stress that we have
used high quality data to fix the size of the wave function and that 
because data set $(a)$ corresponds to parallel kinematics
no experimental information on the $R^{TL}$ response has been employed.

Let us now focus on the spectroscopic factors obtained from reduced
cross-sections in data sets $(b)$ and $(c)$.
Fig.~3 shows the reduced cross-sections for $p_{1/2}$ and 
$p_{3/2}$ shells. Left and right panel correspond  respectively
to data sets $(b)$ and $(c)$. 
As in Fig.~1, 
for each curve
a global scale factor has been fitted to the experimental data.
The corresponding scale factors 
and their statistical errors are listed in Table~III.
Similarly to what we saw for set $(a)$,
also for sets $(b)$ and $(c)$ EDAD-type optical potentials give larger 
spectroscopic factors than EDAI-O (see also Fig.~4).

The results for the $p_{3/2}$ shell corresponding
to the Saclay experiment (left panel) include the contribution of the
$(5/2^+,1/2^+)$ doublet. 
We have verified that
the change in the shape of the responses or reduced cross-section
after inclusion of the doublet is small. The main effect 
of its inclusion is a decrease
of the deduced spectroscopic factor for the $p_{3/2}$ shell of the 
order of $10\%$. In ref.~\cite{Chin91} 
the contribution
of this doublet was subtracted from the experimental data 
with a procedure based on a nonrelativistic
formalism. We have chosen to use the 
uncorrected data from ref.~\cite{Chin-PhD},
and include the contribution from the doublet in our
theoretical calculation. The $s-d$ content has been determined through fits to
data set $(a)$ for this state~\cite{Leus94}.
The values of the spectroscopic factors are
$S_{1/2+}=0.034(2)$ (RCC1), $0.034(2)$ (RCC2),
$S_{5/2+}=0.086(5)$ (RCC1), $0.088(5)$ (RCC2)
(with a $\chi^2/dof$ of the order of 0.5). 
 
Let us now discuss the results corresponding to data set $(b)$ in
left panel of Fig.~3. As shown, the calculations reproduce in general
the experimental data for both shells with the scale factors listed in
Table~III.
Although the various approximations
give similar results, we note that the RCC1 (thin solid line) 
reduced cross-sections for the $p_{1/2}$ shell are less symmetrical around
$p_m=0$, a behavior that is not favored by the data.
For this data set $(b)$ all the calculations, except RCC1
for the $p_{1/2}$ shell, reproduce well the asymmetry
of the reduced cross-section. We will return to this point   
when talking about the $TL$ observables in next section.
Finally, it is important to remark  that the spectroscopic factors
obtained from the  data set $(b)$ (Table~III)
agree, within statistical errors, with  those obtained from data
set $(a)$
taking into account the systematic error of both
experiments: around 5.4\% for data set $(a)$~\cite{Leus94} and 6.3\% for
data set $(b)$~\cite{Chin-PhD}.

Concerning set $(c)$~\cite{Spal93} the data on
reduced cross-sections in right panel of Fig.~3 have been obtained
from the differential cross-sections and detailed kinematics setup  
in Appendices A and D of ref.~\cite{Spal-Md}
(the systematic error for data set $(c)$ reduced
cross-section is 6\%~\cite{Spal93}).
For the $p_{1/2}$ shell, the reduced cross-section is
well reproduced by both relativistic and projected calculations,
except  in the case of the RCC1 calculation
(thin solid line) that underestimates the
data for negative missing momentum values.
This is consistent with the results previously discussed 
for data set $(b)$. For this shell, the spectroscopic factors that
fit data set $(c)$ are larger than the ones derived from
data sets $(a)$ or $(b)$, but they are all compatible
within statistical errors.
In the case of the $p_{3/2}$ shell, although the shape of the
cross-section is well reproduced by the various calculations, the
situation on the spectroscopic factors is clearly different
(see Table~III). With EDAI-O optical potential the values
of the spectroscopic factors that fit the $p_{3/2}$ data on
reduced cross-sections in set ($c$) are 
$25-30\%$ larger than the ones obtained  from 
data set $(a)$. 
These scale factors are also larger than the ones obtained from
 data set $(b)$, but in this case the discrepancy is of the order of
15\%, which is comparable to the combined 
systematic and statistical error for these values. EDAD-type potentials not
only give larger spectroscopic factors but also give, on average, better
agreement between $p_{3/2}$ spectroscopic factors of the three different
sets ($a$), ($b$) and ($c$). This is seen in detail in Tables I and III and is
further illustrated in Fig.~4.

In summary,  the shapes of the reduced
cross-sections are well described 
by  all the RCC2 calculations and data sets, what makes us conclude that
 we can  rely on the
spectroscopic factors derived with EDAD-1 and NLSH-P potentials. 
Thus the differences
in the $p_{3/2}$ spectroscopic  factors (see Fig.~4) obtained with the same 
ingredients (wave functions,
operators and optical potentials) may be attributed 
either to a global scale
variation among the three experiments for the $p_{3/2}$ shell, or to
limitations of the theory.
Coupled channel contributions or MEC could possibly make a different 
effect for the three kinematics analyzed in this work.

\subsection{Response Functions and longitudinal-transverse Asymmetry}

In this section we present results for the response functions and
asymmetries and compare them to the data in sets
$(b)$ and $(c)$ measured at Saclay~\cite{Chin91}
and  NIKHEF~\cite{Spal93}, respectively. 
As already mentioned, these two experiments 
were performed under $|\vec{q}|-\omega$ constant kinematics so that 
the $TL$ response and asymmetry ($R^{TL},A_{TL}$)
can be obtained from the cross-sections measured at 
$\phi_F=0^o$ and $\phi_F=180^o$ with the other variables 
($\omega$, $Q^2$, $E_m$, $p_m$) held constant. Moreover, the response
functions $\displaystyle R^L+\frac{q^2}{2Q^2}R^{TT}$ and $R^T$ 
were also determined for data set $(c)$~\cite{Spal93}.

Figs.~5 and 6 show respectively $R^{TL}$ and 
$A_{TL}$ for $p_{1/2}$ (upper panels) and $p_{3/2}$
(lower panels)  corresponding to set $(b)$  (left panels) and 
set $(c)$ (right panels). In each panel we present four
curves with the same conventions as in previous
figures: RCC1 (thin solid), RCC2
(thick solid), PCC1 (thin-dashed)  and PCC2 (thick-dashed).
Each $R^{TL}$ curve
is scaled with the corresponding spectroscopic factor 
quoted in Table~III. As it was also the case for reduced cross-sections, there
are no appreciable differences in the shapes of curves obtained with the
different types of optical potentials.
Obviously, the asymmetry $A_{TL}$ is independent on the value of the
spectroscopic factor. The results for the
$p_{3/2}$ shell in bottom-left panel of Figs. 5 and 6 include the
contribution 
of the ($5/2^+$,$1/2^+$) doublet as explained for set ($b$)
in previous section. The asymmetry
$A_{TL}$ was not produced by the Saclay experiment (set$(b)$), 
but we have deduced
$A_{TL}$ from the data using the $R^{TL}$ values as well as
the cross-section data in~\cite{Chin-PhD}. 

In Fig.~5 we also show by dotted lines the nonrelativistic results of
ref.~\cite{Spal93}. For comparison to previous studies 
in refs.~\cite{Chin91,Spal93}, we quote
in Table~IV the factor required to scale
the theoretical predictions to the $R^{TL}$ data, additional to the
factors in Table~III. A value of one in this table indicates that the same
spectroscopic factor fits {\em both} the reduced cross-section and
$R^{TL}$, {\it i.e.,} indicates that the $TL$ strength is consistently 
predicted by the theory.

Let us first  discuss  the comparison between theory and
experiment for data set $(b)$.
From the results shown in Figs.~5 and 6, it is clear that
the effects of the negative-energy components show up
 more in $R^{TL}$ and
$A_{TL}$ than in the cross-sections (Fig.~3).
In the case of the $p_{1/2}$ shell (left-top panel of Fig.~5),
the RCC2 calculation
agrees with experimental data within statistical errors,
while PCC1 and PCC2 results for $R^{TL}$ (dashed lines)
lie about a 30-50\% below the data, and the RCC1
calculation (thin solid line) overestimates 
the $R^{TL}$ response by around 20\% (see Table~IV).
In the case of the $p_{3/2}$ orbit (left-bottom panel), all
the approximations predict similar curves:
The projected results are  much closer to fully relativistic ones
than for the $p_{1/2}$ shell. Overall, the
fully relativistic calculations seem to be favored by the data.
The fact that in this shell the variation introduced by the 
negative energy components is much
smaller than for the $p_{1/2}$ shell explains why
the difference between RCC1 and RCC2 results is smaller for the $p_{3/2}$
than for the $p_{1/2}$ shell. These results agree with
the conclusion reached from RPWIA calculations in ref.~\cite{Cab98b}
about the behavior of $j=\ell\pm 1/2$ spin-orbit partners which was also
corroborated in RDWIA calculations at high $|Q^2|$~\cite{Udi99}.

With regards to the $TL$ observable, independent on the spectroscopic factor,
we may conclude that
for $p_{1/2}$ shell, $A_{TL}$ is best reproduced by RCC2 results, while for
$p_{3/2}$ shell the four 
theoretical results are very close together, and the experimental data 
agree with all of them.

In the right panels of Figs.~5 and 6 we see  the results corresponding
to data set $(c)$.
Most of the comments on data set $(b)$ apply also here,
though the data are somewhat more scattered and have larger error bars.
In the case of the $p_{1/2}$ shell, 
PCC1 and PCC2 results are very similar and lie below the data;
among the fully relativistic calculation, the
RCC2 result reproduces the data within statistical errors, while RCC1
overestimates them by a 35\%. 
In the case of the $p_{3/2}$ shell (bottom-right panel), all the
calculations underestimate the experimental $TL$ response by around 17-28\%,
except RCC1 for which the ``additional'' factor in Table~IV is
compatible with one within statistical errors.

In Fig.~7 we show the results for the responses $R^L+v_{TT}/v_L R^{TT}$
(top panels) 
and $R^T$ (bottom panels) for the $p_{1/2}$ and $p_{3/2}$ shells
compared to the data from NIKHEF~\cite{Spal93}. 
Each curve is scaled with the spectroscopic factors quoted
in Table~III.
Notice that these responses are rather insensitive to 
dynamical enhancement of lower components. This is consistent with
the behavior observed in Fig.~1 and also with results of RPWIA 
calculations~\cite{Cab98b}.
The results in Fig.~7 indicate that
the separated  responses are in general well 
reproduced by the relativistic as well as by the projected calculations
for both shells,
exception made of the data point at the lowest missing momenta where,
as indicated by the large error bars, the $L/T$ separation is more problematic.

Summarizing, for the $p_{1/2}$ shell 
the RCC2 results agree well with all
observables and data sets, while RCC1 (projected) calculations
show a too large (small)
$R^{TL}$ and $A_{TL}$. For the $p_{3/2}$ shell the theoretical calculations
lie much closer together, and generally agree
with all data sets and observables, 
except for $R^{TL}$ and $A_{TL}$ of data set $(c)$.
Although the $R^{TL}$, $A_{TL}$ data on $p_{3/2}$ in set ($c$) lie
higher than theoretical calculations,
they are almost compatible with RCC1 and
RCC2 calculations within statistical errors.
This situation is quite different from the one  found in ref.~\cite{Spal93}, which
is also shown for comparison in Fig.~5. The dotted lines in this figure show
the nonrelativistic results of ref.~\cite{Spal93} that were obtained with
nonrelativistic spectroscopic factors (0.61(3) for $p_{1/2}$ and 0.53(3) for $p_{3/2}$)
 and
standard (Woods-Saxon type) nonrelativistic optical potentials and bound
wave functions. The latter were also fitted to Leuschner data~\cite{Leus94}.

\subsection{Further Discussion on $TL$ observables}
 
In this section we focus in Fig.~5 comparing our results to
previous nonrelativistic ones. The dotted curves -representing the
nonrelativistic calculations by Spaltro {\it et. al.}~\cite{Spal93}-
clearly underestimate $TL$ responses for all shell and data sets. 
The deviation from data is larger for $p_{3/2}$ shell, particularly
 in data set ($c$), where the dotted curve gives roughly one
half of the experimental $TL$ response. Why is it that relativistic
results in this figure are so much closer to data that nonrelativistic ones?.

We have examined in detail effects due to the various aspects 
that are relevant in comparing relativistic to nonrelativistic
results. The effects of Darwin term are already taken into account as they
basically affect the spectroscopic factors. The effects of the negative
energy components, as already mentioned, are very small for $R^L$, $R^T$
responses in all data sets ($a$), ($b$) and ($c$) but, as seen in Fig.~5 and
Table~IV, they are important for $TL$ observables in data sets ($b$) and ($c$),
particularly for the $p_{1/2}$ shell. We are then left to consider the
effect of truncation of the current operator (TCO). TCO produces also a
negligible effect at the kinematics of data set ($a$), but it is more
important at the kinematics of data sets ($b$) and ($c$). This again affects
more the $TL$ responses and asymmetries where it may represent up to
 a $15\%$ effect (see also ref.~\cite{MGP01}).
Thus for $p_{1/2}$ shell, TCO roughly
explains the difference at the maxima between dotted curves and
the curves obtained with projected calculations. However for $p_{3/2}$ shell
TCO explains only a small fraction of the difference between dotted curves and
results of projected calculations. The largest fraction of this difference is
due to the use of a too small spectroscopic factor (see
ref.~\cite{Spal93}) that was taken from ref.~\cite{Leus94} and that by no means
fits the data on reduced cross-sections in set ($c$). 
As seen in Tables I, III and
Fig.~4 the spectroscopic factor deduced from reduced cross-sections in data
set ($c$) is $25-30\%$ larger than that from data set ($a$).

The message from this is, not only that relativistic effects are important
in perpendicular kinematics at low $|Q^2|$, but also that a careful analysis
of all pieces of information has to be done to get a consistent picture of
the three different sets of data. Since $R^{TL}$ responses are known to be
sensitive not only to relativistic effects but also to exchange currents,
or other possible many-body effects, it is important to establish a clear
framework that allows to look for the proper magnitude of such effects.

Indeed if we compare our results to data for the $A_{TL}$ observable that
is free from spectroscopic factor ambiguities, we find that all data are well 
reproduced with the standard CC2 current operator, except the $p_{3/2}$ data
in set ($c$) which are only larger than theory by a factor $\sim 1.17$. This
is to be compared to the $2.05$ factor that one could expect from 
ref.~\cite{Spal93}.

\section{Summary and Conclusions}

In summary we find that the fully relativistic treatment 
improves substantially the 
description of reduced cross-sections and individual responses of all
three sets of data on $^{16}O(e,e'p)$ at low $|Q^2|$. Although predictions
from CC1 and CC2 current operators are rather close in most cases, data
seem to favor the CC2 current operator. Therefore our remarks here will
focus mainly on
results with CC2 and with the improved NLSH-P bound nucleon wave
functions, that have the correct r.m.s. radius. Using the most complete
set of data on reduced cross-sections in parallel kinematics of
Leuschner {\em et al.}~\cite{Leus94} (set $(a)$) 
we obtain spectroscopic factors
ranging from 0.58 to 0.64 for $p_{1/2}$ and from 0.45 to 0.55
for $p_{3/2}$, depending on whether we use A-independent
(EDAI-O) or A-dependent (EDAD-1,2,3) optical potentials. In
$^{16}$O, the latter potentials produce a larger Darwin effect, thus
larger spectroscopic factors. Compared to the cases studied in previous works
on $^{40}Ca$ and $^{208}$Pb, the determination of spectroscopic factors in
$^{16}$O with the relativistic approach is different in several
respects. In the former cases, the standard NLSH wave functions were 
found to reproduce well the shapes of reduced cross-sections and the only fitted
parameter was the spectroscopic factor. The latter was practically independent on the
optical potential used and was $\sim 0.7$ for the levels just below the Fermi 
level.
On theoretical grounds smaller spectroscopic factors for $^{16}$O are 
 expected. In particular, from shell model Monte Carlo calculations on 
$^{16}$O\cite{Otsuka}, one may expect $S_\alpha\sim 0.5$ though
other theories predict somewhat larger values~\cite{VN94,MD97}.
Larger spectroscopic factors are obtained from Spaltro {\em et al.} data \cite{Spal93}
on reduced cross-sections in perpendicular kinematics (set ($c$)), while
Chinitz's {\em et al.} data \cite{Chin91} also in perpendicular kinematics
(set ($b$)) give similar spectroscopic factors than set ($a$). As one
can see in Fig.~4, within error bars, spectroscopic 
factors derived from all data sets with EDAD-1 are compatible with
each other. To overcome the uncertainty due to the optical potential (see
also Fig.~4) one would need to fit the relativistic potential to both
elastic and inelastic proton scattering data from $^{16}O$ in a manner similar
to what has been done for nonrelativistic potentials~\cite{KellyEEI}.
The analyses of individual responses is practically independent on
the optical potential, once they are scaled by the corresponding spectroscopic
factors.

There is a long standing controversy surrounding the $TL$ data for the
$p_{1/2}$ and $p_{3/2}$ shells measured at Saclay~\cite{Chin91} and
NIKHEF~\cite{Spal93}. We have therefore paid particular attention to $TL$
responses and asymmetries and we conclude that there is not
a fundamental
inconsistency. Even at the low $|Q^2|$ values considered here,
the $TL$ response is much more sensitive than $L$ and $T$ 
responses to
relativistic effects,  in particular to the dynamical enhancement of
the lower components. The role played by the latter is appreciated
comparing fully relativistic results (RCC2 or RCC1) to those obtained
using wave functions projected on the positive energy sector (PCC2 or
PCC1). RCC2 results agree well with experimental $TL$ responses on $p_{1/2}$
(as well as with $TL$ asymmetries) which are underestimated by PCC2 and
overestimated by RCC1, because CC1 current operator overemphasizes the
role of 
negative energy components. The overall agreement with data on $TL$
responses and asymmetries from set $(b)$  and
set $(c)$  is quite satisfactory, with the exception of data
on $p_{3/2}$ shell from set $(c)$, but even in this case theory is much
closer to experiment than previously found in ref.~\cite{Spal93}.
 In particular,
the large difference between data on $TL$ responses from the two different
sets is well accounted for by the present analyses. 
This is in contrast with the situation depicted in ref.~\cite{Spal93},
which is represented by dotted lines in Fig.~5.

In short, the puzzle of the large discrepancy in the TL-response obtained in
Saclay~\cite{Chin91} and NIKHEF~\cite{Spal93}, and the ``additional''
TL-strength found in both experiments is, to a large extent, explained by the
effect of the negative energy components in the wave functions --a dynamical
relativistic effect that may not have been expected at low transfer and missing
momentum.

The large general mismatch of  data set $(c)$
on $p_{3/2}$ shell seems to point to a normalization problem which would
require experimental verification. Our analyses indicate that the problem
is not so much connected to the $TL$ response, but rather to the
normalization used.
Nevertheless, since meson
exchange currents and particularly isobar currents are claimed to affect
more the $p_{3/2}$  than the $p_{1/2}$ orbitals~\cite{Ryck99}, 
it would be interesting to see whether our fully 
relativistic calculation extended to include the isobar 
and other meson exchange
effects would  lead to better agreement with $TL$ $p_{3/2}$ data from
set $(c)$. It will also be interesting to see how relativistic and nonrelativistic 
approaches compare to new data expected from future
experiments that have been approved to measure 
reduced cross-sections and TL-responses in $^{16}O$ with 
unprecedented precision at Jefferson Lab.

This work was partially supported under Contracts  No. 
PB/98-1111, PB/98-0676, PB/96-0604
and by the Junta de Andaluc\'{\i}a (Spain). J.R.V. and A.E. acknowledge
support from  doctoral fellowships of the Consejer\'{\i}a de Educaci\'on
 de la Comunidad de Madrid and Ministerio de Educaci\'on y Cultura
 (Spain), respectively.

\begin{table}
\begin{center}
\begin{tabular}{|c|c|cc|cc|cc|cc|}
  & &\multicolumn{4}{c|}{NLSH} 
& \multicolumn{4}{c|}{NLSH-P}\\ 
\hline 
 &  &  \multicolumn{2}{c|}{$p_{1/2}$}  & \multicolumn{2}{c|}{$p_{3/2}$} &
     \multicolumn{2}{c|}{$p_{1/2}$}  & \multicolumn{2}{c|}{$p_{3/2}$} \\ 
& & CC1    & CC2   & CC1   & CC2   & CC1   & CC2   &     CC1  
& \multicolumn{1}{c|}{CC2}   \\ \hline \hline
& & \multicolumn{8}{c|}{RELATIVISTIC} \\ \hline
EDAI-O &$S_\alpha$ & 0.58(1) & 0.64(2) & 0.45(3)
& 0.49(3)  & 0.54(1) & 0.58(1) & 
0.43(1) & \multicolumn{1}{c|}{0.45(1)}\\ 
  & $\chi^2/dof$ & 6.6 & 4.5 & 25.3 & 15.7 & 1.3 & 1.3 & 2.7 &
\multicolumn{1}{c|}{3.5} \\ \hline
EDAD-1 &$S_\alpha$ & 0.63(4) & 0.72(2) & 0.56(3) & 0.62(2) 
& 0.58(1) & 0.64(1) & 
0.52(1) & \multicolumn{1}{c|}{0.55(2)}\\ 
  & $\chi^2/dof$ & 9.6 & 3.7 & 15 & 7.2 & 1.2 & 1.2 & 1.2 &
  \multicolumn{1}{c|}{4.8} \\  \hline
EDAD-2 &$S_\alpha$ & 0.61(4) & 0.69(3) & 0.53(3) & 0.59(2) 
& 0.56(1) & 0.62(1) & 
0.50(1) & \multicolumn{1}{c|}{0.52(1)}\\ 
  & $\chi^2/dof$ & 10 & 2.6 & 18 & 9.2 & 1.4 & 1.1 & 1.7 &
  \multicolumn{1}{c|}{4.1} \\  \hline \hline 
  & & \multicolumn{8}{c|}{PROJECTED}  \\ \hline
EDAI-O & $S_\alpha$ & 0.65(2) & 0.66(2) & 0.51(3) & 0.52(3)
& 0.58(1) & 0.59(1) &
0.47(1) & \multicolumn{1}{c|}{0.46(1)} \\  
  & $\chi^2/dof$ & 4.5 & 3.2 & 16.9 & 13.3 & 1.2 & 1.6 & 3.6 &
\multicolumn{1}{c|}{4.3} \\ \hline
EDAD-1&$S_\alpha$ & 0.72(3) & 0.74(2) & 0.64(3) & 0.64(2) & 0.64(1) & 0.65 (2) & 0.57(2) 
 & \multicolumn{1}{c|}{0.56(2)}\\
 & $\chi^2/dof$ & 4.0  & 2.6  & 7.6  & 5.9  & 1.4 &  1.6 & 4.4
 & \multicolumn{1}{c|}{6.3} \\ \hline
EDAD-2&$S_\alpha$ & 0.69(3) & 0.71(2) & 0.61(3) & 0.61(3) & 0.62(1) & 0.63(1) & 0.55(2) 
 & \multicolumn{1}{c|}{0.54(2)}\\ 
 & $\chi^2/dof$ &4.6  & 3.1  & 9.7  & 7.4  & 1.3  & 1.5  & 3.7
 & \multicolumn{1}{c|}{5.3} \\
\end{tabular}
\caption[Table I]{Spectroscopic factors derived from
Leuschner's experimental reduced cross
sections in ref.~\protect\cite{Leus94} (data set $(a)$) using NLSH
and NLSH-P relativistic bound nucleon wave functions, and
EDAI-O, EDAD-1 and EDAD-2 relativistic optical potential parameterizations
(see text). Results with EDAD-3 are almost identical to the ones with EDAD-1.
The numbers within
parentheses indicate the statistical error.}
\end{center}
\end{table}

\begin{table}
\begin{center}
\begin{tabular}{|c|ccc|ccc|} 
 & \multicolumn{3}{|c|}{$p_{1/2}$} &  \multicolumn{3}{c|}{$p_{3/2}$}\\  \hline
        & $b.e.$ (MeV)& $r.m.s.r.$-$r$ (fm) & $r.m.s.r.$-$p$ (MeV) &
         $b.e.$ (MeV) & $r.m.s.r.$-$r$ (fm) & 
\multicolumn{1}{c|}{$r.m.s.r.$-$p$ (MeV)} \\  \hline
NLSH    & 11.4 & 2.838 & 175.7 &18.8  & 2.679 & \multicolumn{1}{c|}{185.2}   \\ 
NLSH-P  & 12.1 & 3.043 & 170.6 &18.4  & 2.907 & \multicolumn{1}{c|}{173.6} \\  
\end{tabular} 
\caption[Table II]{Comparison of binding energies and {\em r.m.s.r.} radius in
$p-$ and $r-$ space
for the  wave functions
NLSH \protect\cite{Shar93}  and 
NLSH-P. The contribution from the negative energy
components to the norm of the wave function is about 2\% in all cases.}  
\end{center} 
\end{table} 

\begin{table}
\begin{center}
\begin{tabular}{|c|cc|cc|cc|cc|}
& \multicolumn{4}{c|}{Set $(b)$ Chinitz {\em et al.}\protect\cite{Chin91}}&
\multicolumn{4}{c|}{Set $(c)$ Spaltro {\em et al.}\protect\cite{Spal93}}\\ 
\hline 
   &  \multicolumn{2}{c|}{$p_{1/2}$}  & \multicolumn{2}{c|}{$p_{3/2}$} &
     \multicolumn{2}{c|}{$p_{1/2}$}  & \multicolumn{2}{c|}{$p_{3/2}$}\\ 
 & CC1    & CC2   & CC1   & CC2   & CC1   & CC2   &     CC1 
& \multicolumn{1}{c|}{CC2}   \\ \hline
EDAI-O (R)
& 0.54(4)  & 0.56(3) & 
0.49(2) & 0.51(2) & 0.57(3) & 0.61(2) & 0.56(1) & 
\multicolumn{1}{c|}{0.59(2)}\\ 
EDAI-O (P) 
& 0.59(4) & 0.56(4) &
0.53(4) & 0.53(3) & 0.66(2) & 0.63(2) & 0.61(1) & 
\multicolumn{1}{c|}{0.61(2)} \\    \hline
EDAD-1 (R)
& 0.59(4)  & 0.61(3) & 
0.53(3) & 0.55(3) & 0.68(4) & 0.72(2) & 0.62(2) & 
\multicolumn{1}{c|}{0.67(2)}\\ 
EDAD-1 (P) 
& 0.65(4) & 0.62(3) &
0.57(5) & 0.57(4) & 0.79(3) & 0.74(3) & 0.69(3) & 
\multicolumn{1}{c|}{0.69(3)} \\ 
\end{tabular}
\caption[Table 2]{Spectroscopic factors derived from
two different sets of data on experimental reduced cross
sections from  the full relativistic approach (R) and
from the projected one (P). The nomenclature used is the same as in Table~I.
The numbers within
parentheses show the statistical error only. All  results correspond
to the NLSH-P bound wave function.}
\end{center}
\end{table}

\begin{table}
\begin{center}
\begin{tabular}{|c|cccc|cccc|} &
\multicolumn{4}{c|}{Set $(b)$}& \multicolumn{4}{c|}{Set $(c)$} \\ \hline
 & \multicolumn{2}{c}{$p_{1/2}$}  &\multicolumn{2}{c|}{$p_{3/2}$} &
   \multicolumn{2}{c}{$p_{1/2}$}  & \multicolumn{2}{c|}{$p_{3/2}$} \\
 & $N_{TL}$ & $\chi^2/dof$ & $N_{TL}$ & $\chi^2/dof$ & $N_{TL}$ & $\chi^2/dof$ & 
$N_{TL}$ &
\multicolumn{1}{c|}{$\chi^2/dof$} \\ 
RCC1 & 0.83(10) & 0.65  & 0.95(17) &  5.3& 0.63(10) & 1.1 & 1.09(12)& 
\multicolumn{1}{c|}{3.2} \\
PCC1 & 1.32(42) & 4.5  & 1.14(22) & 5.8 &1.15(17)  & 1.1  & 1.28(12) &
\multicolumn{1}{c|}{ 2.5}\\ 
\hline
RCC2 & 1.14(12)&0.49  & 1.02(15)& 3.4 & 0.90(13)&1.0 & 1.17(12)& 
\multicolumn{1}{c|}{2.5} \\ 
PCC2 &  1.48(32)& 2.2  & 1.11(17)& 3.8 & 1.26(19)&1.0 & 1.28(11)&
\multicolumn{1}{c|}{2.1} \\ \hline
NR &1.56(12)   & --- & 1.66(9)& ---  & 1.50(12)& ---  & 2.05(10)&
\multicolumn{1}{c|}{---} \\
\end{tabular}
\caption[Table IV]{Extra  scale factor $N_{TL}$ needed to fit
the experimental $R^{TL}$ response. These factors would
multiply those in Table~III to scale theory to experiment
on $R^{TL}$.
A value of one  indicates
that no extra enhancement or quenching of the response is found.
The numbers in parentheses show     the statistical error only.
The quality of the fit ($\chi^2/dof$) is also quoted
in every case. NR corresponds to the nonrelativistic analysis of
ref.~\protect\cite{Spal93}. These numbers correspond
to EDAI-O potential. Very similar numbers are obtained with EDAD-1,
EDAD-2 or
EDAD-3.} 
\end{center}
\end{table}

\begin{figure} 
\begin{center}
\leavevmode 
\mbox{\epsfig{file=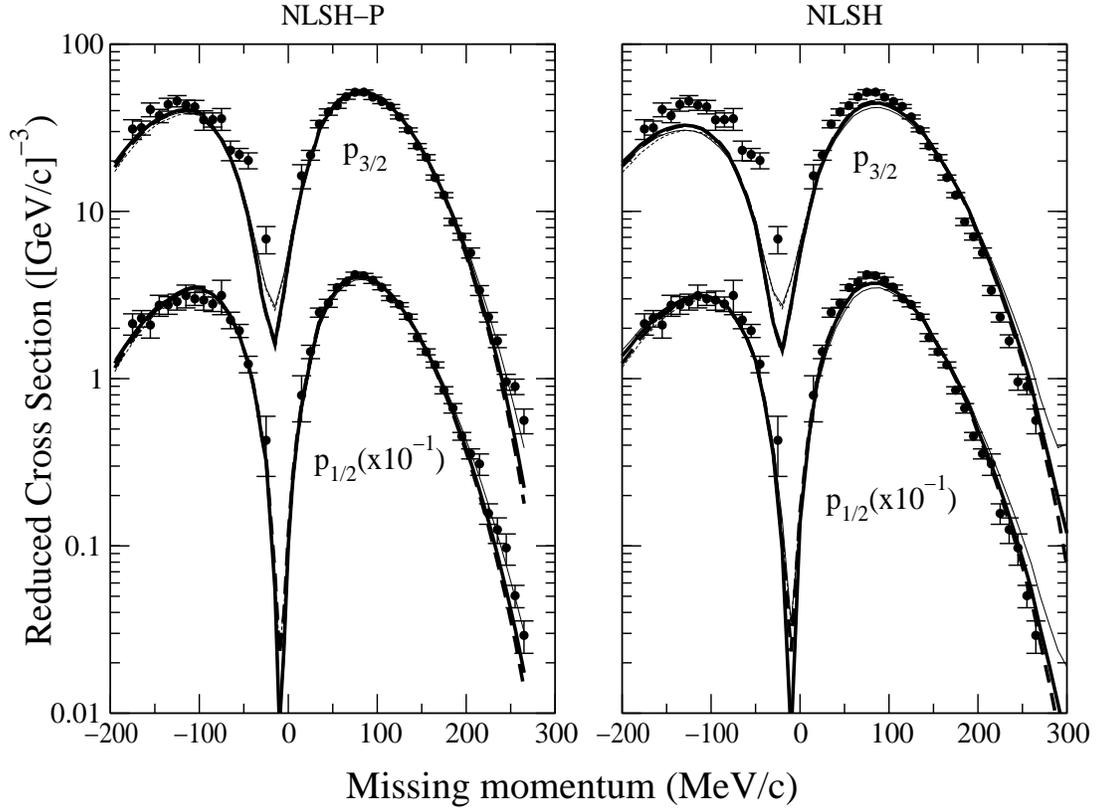,height=0.55\textheight}}
\end{center} 
\caption{Reduced cross-sections for proton knockout  from
 $1p_{1/2}$  and $1p_{3/2}$ orbits in $^{16}O$
versus missing momentum $p_m$  corresponding to
the experiment performed by Leuschner
 {\em et al.}~\protect\cite{Leus94} (set $(a))$. 
The bound relativistic proton wave function has been obtained with the 
NLSH (right panel) and NLSH-P (left panel) parameterization.
Theoretical results shown correspond to a fully relativistic 
calculation using the Coulomb gauge and  current  
operators RCC1 (thin solid line)
and RCC2 (thick solid line). The optical potential used
is EDAI-O from ref.~\protect\cite{CHCM93}. Also shown are the results after  
projecting the bound and scattered proton wave 
functions over positive-energy  
states: PCC1 (thin dashed line), PCC2 (thick dashed line). 
EMA-noSV results (not shown) are 
practically identical to PCC2 ones. 
Each curve is scaled by the corresponding spectroscopic factor
in Table~I.
} 
\end{figure} 
 
\begin{figure} 
\begin{center}
\leavevmode 
\mbox{\epsfig{file=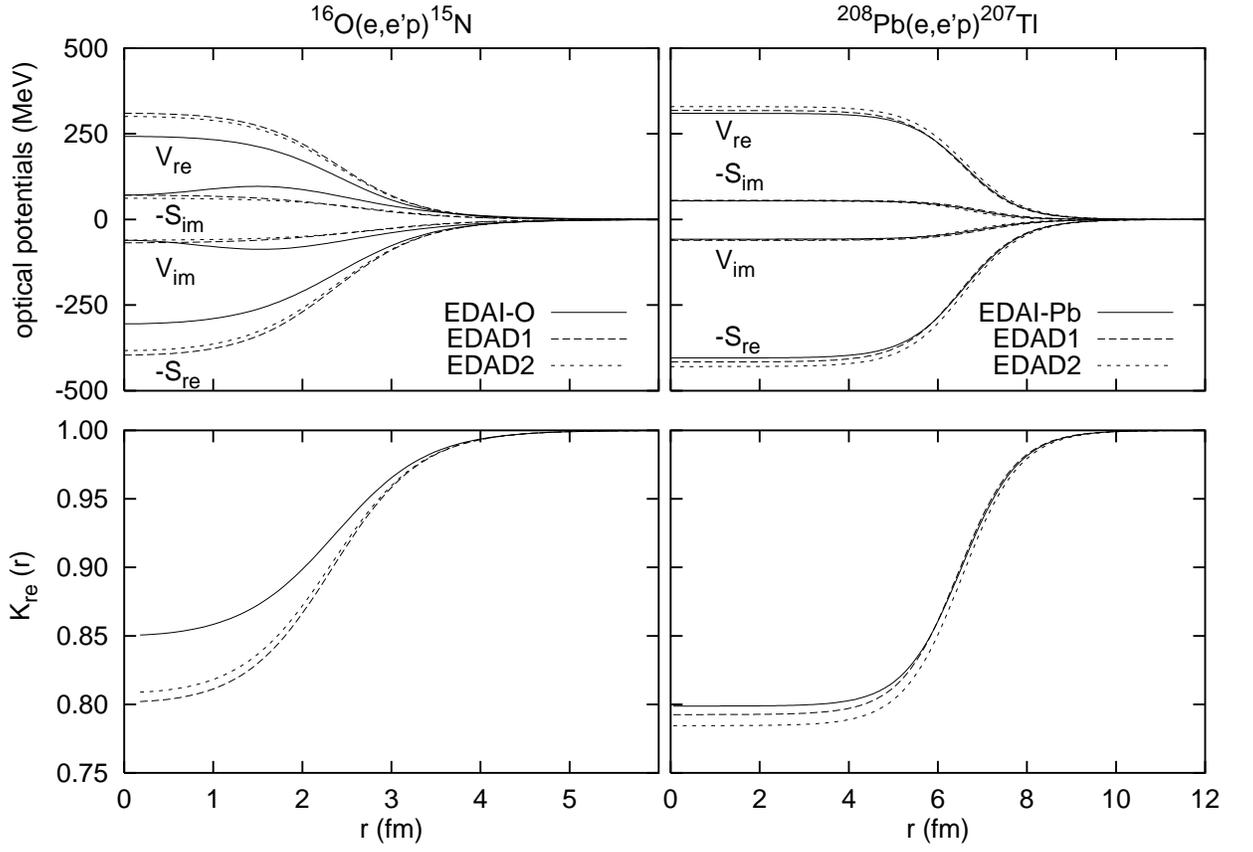,height=0.55\textheight}}
\end{center} 
\caption{Real and imaginary part of the optical potentials (upper panels) and 
real part of the Darwin
factor (lower panels) (the imaginary part is negligible) for $^{16}O$ (left panels) and
$^{208}Pb$ (right panels).}
\end{figure}

\begin{figure} 
\begin{center}
\leavevmode 
\mbox{\epsfig{file=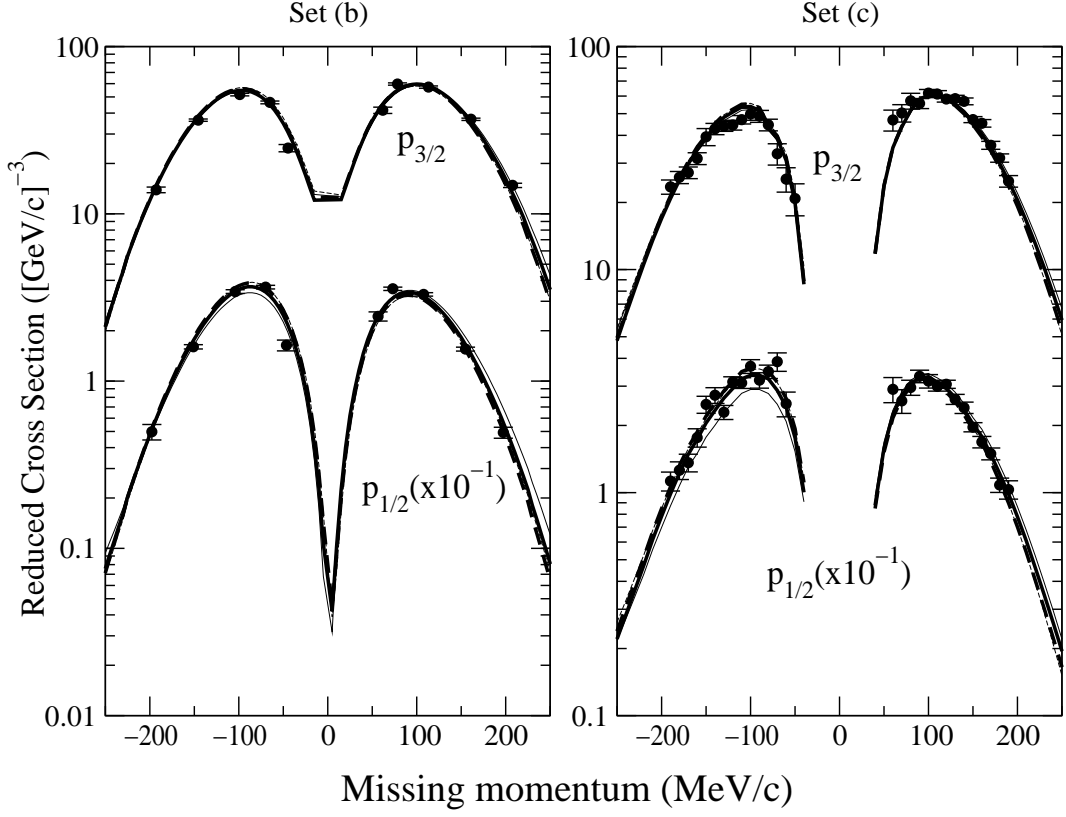,height=0.55\textheight}}
\end{center} 
\caption{Same as Fig. 1  for the experiments performed by
 Chinitz {\em et al.}\protect\cite{Chin91} 
(left panel, set $(b)$) and by 
 Spaltro 
{\em et al.}\protect\cite{Spal93} (right panel, set $(c)$). In 
all the cases the NLSH-P
relativistic bound proton wave function and EDAI-O optical potential
 has been used. 
For  $p_{3/2}$ shell in set $(b)$
 the contribution from the nearby $5/2^+$
and $1/2^+$ states has been taken into account (see text).
Each curve is scaled by the corresponding spectroscopic factor
in Table~III.}
\end{figure}

\begin{figure}
\begin{center} 
\mbox{\epsfig{file=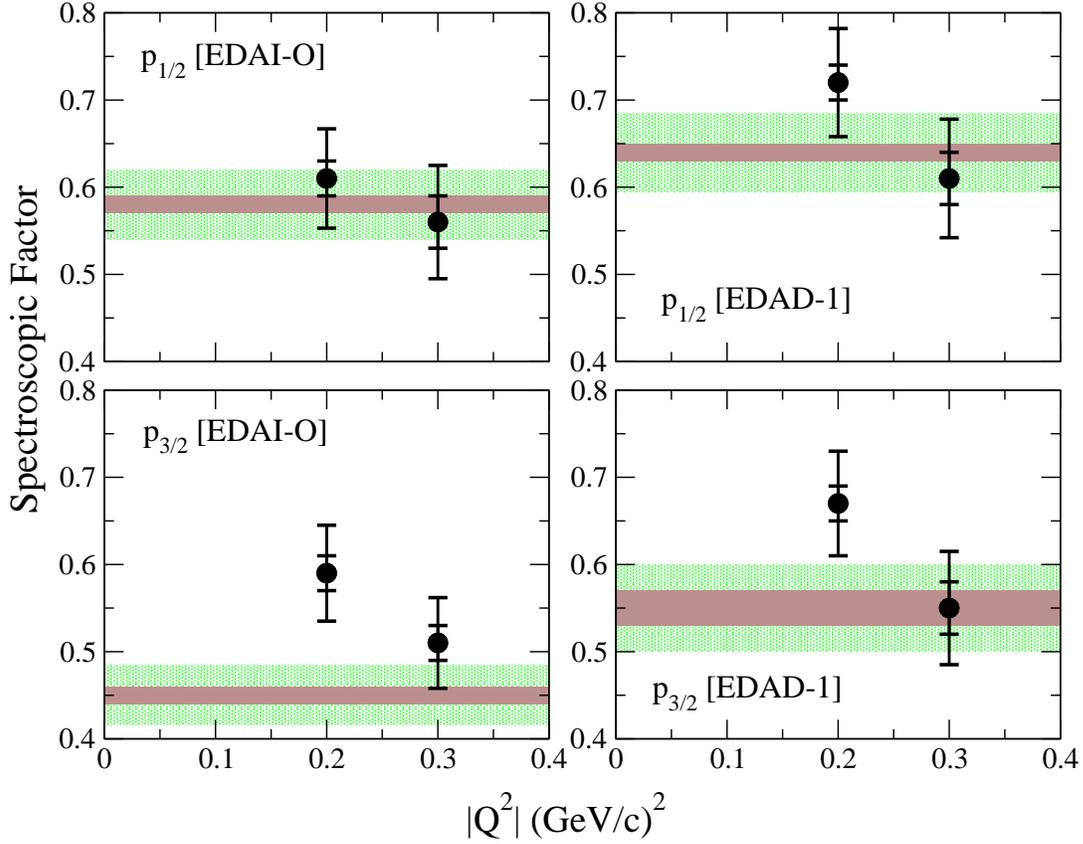,width=\textwidth}} 
\end{center} 
\caption{Spectroscopic factors derived within the fully relativistic
approach from the low-$Q^2$ data
discussed in this work with NLSH-P wave function, CC2 current operator
and EDAI-O (left) or EDAD-1 (right) optical potentials.
 The inner error bars include statistical
errors only, the outer one includes also the additional
systematic error in the reduced cross-sections for each experiment.
The bands covering the whole $|Q^2|$ 
range corresponds to 
 the value obtained from the data set $(a)$
\protect\cite{Leus94}, while the dots at $|Q^2|=0.2$ (GeV/c)$^2$ and
0.3 (GeV/c)$^2$ correspond to the data set $(c)$
\protect\cite{Spal93} and set $(b)$ \protect\cite{Chin91},
respectively.}
\end{figure} 

\begin{figure} 
\begin{center}   
\mbox{\epsfig{file=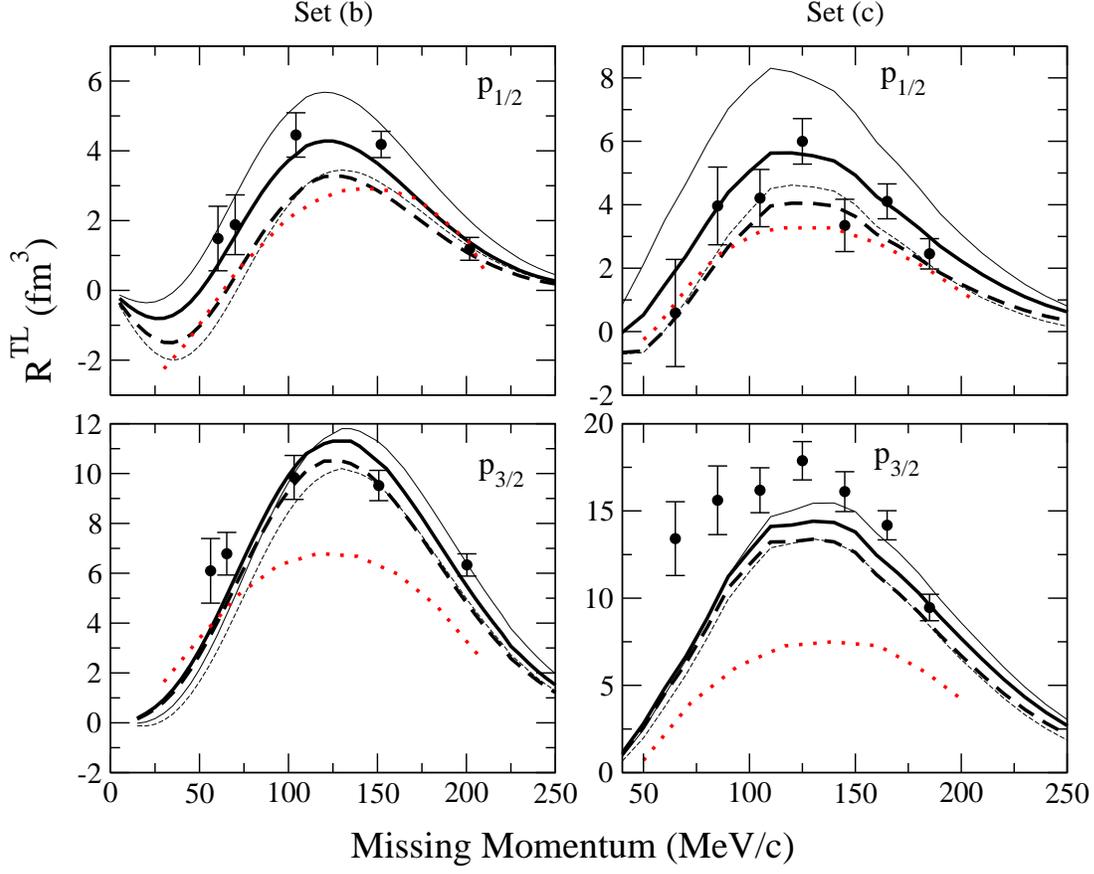,height=0.55\textheight}} 
\end{center} 
\caption{ 
Response
$R^{TL}$
 for proton knockout from $^{16}O$ for
 $1p_{1/2}$ (top panels) and $1p_{3/2}$ (bottom panels).
Results and data shown correspond to kinematics of
data set $(b)$~\protect\cite{Chin91} (left) and 
set $(c)$~\protect\cite{Spal93} (right). Line conventions as
in Figs.~1 and 3 (NLSH-P wave function and EDAI-O optical potential). 
The curves have been  scaled by the 
spectroscopic factors in Table~III.
 Additional dotted curves correspond to the
nonrelativistic analyses of ref.~\protect\cite{Spal93}.}
\end{figure} 
\newpage

\begin{figure}
\begin{center} 
\mbox{\epsfig{file=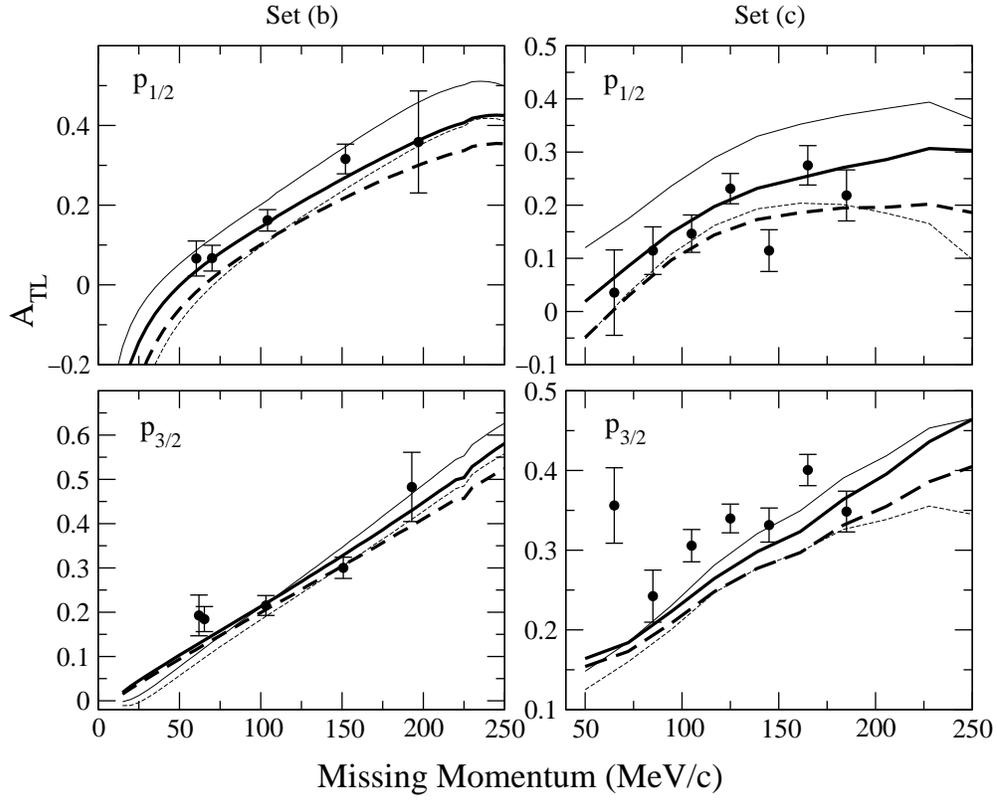,width=\textwidth}} 
\end{center} 
\caption{Same as Fig.~5 for the  
 $A_{TL}$
asymmetry. We recall that this observable is independent on
the spectroscopic factor.}
\end{figure} 

\begin{figure}
\begin{center} 
\mbox{\epsfig{file=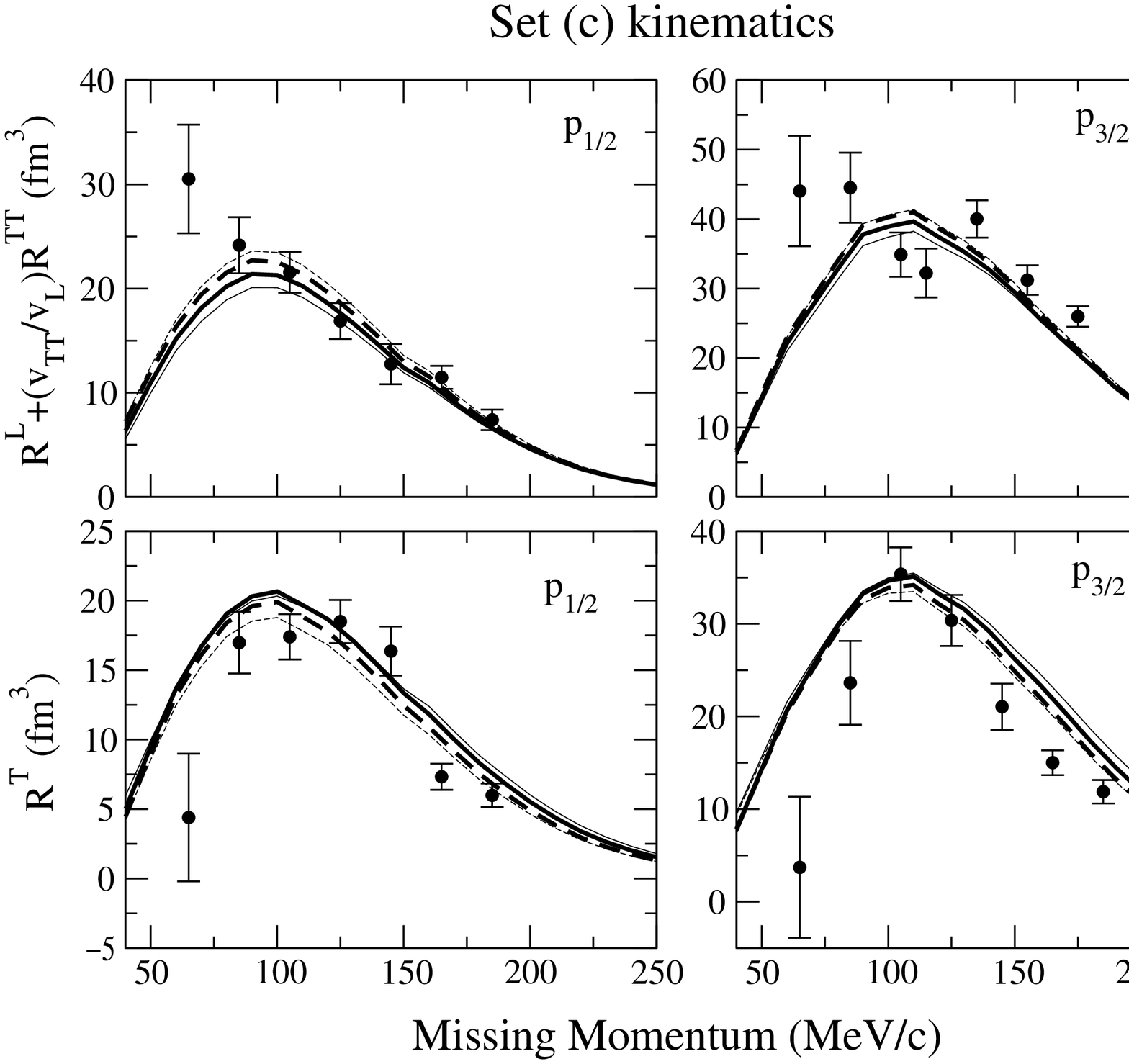,width=\textwidth}} 
\end{center} 
\caption{Response functions $R^{L}+v_{TT}/v_{L}R^{TT}$ and
$R^{T}$ for the kinematics  of  data set $(c)$
\protect\cite{Spal93}. Curves and calculations as in Fig.~3. 
The theoretical results are scaled
with  spectroscopic factors  for this same experiment
in Table~III.}
\end{figure} 
\newpage

\end{document}